# Coexistence of phases and interphase boundaries in BaTiO$_3$

T. A. Liebsch and V. L. Sobolev

South Dakota School of Mines and Technology, Rapid City, SD 57702

**Abstract**

A theoretical analysis of the internal structure of interphase boundaries separating domains of coexisting phases are presented for the perovskite ferroelectric BaTiO$_3$. The temperature dependence of interphase boundary widths and surface energies are calculated and compared with the corresponding parameters for different types of domain walls existing within the ferroelectric phases of BaTiO$_3$.

Barium titanate is one of the most researched perovskite ferroelectrics (FE); however, BaTiO$_3$ and BaTiO$_3$-based compounds continue to attract unabated attention.[1-5] BaTiO$_3$ cooled at ambient pressure undergoes a sequence of first-order phase transitions[1,4,5] from paraelectric phase to FE tetragonal phase (at 404 K) then to FE orthorhombic phase (at 273 K) and finally to FE rhombohedral phase (at 183 K). The vector of spontaneous polarization is directed along the edge (in the tetragonal phase) of the unit cell and then reorients to be along a face diagonal (in the orthorhombic phase) and body diagonal (in the rhombohedral phase). The hysteretic behavior of the temperature dependencies of lattice parameters, dielectric permittivity, birefringence, and spontaneous polarization near all transition points was clearly observed.[6,7] Recently, the monoclinic phase has been observed in the temperature interval between the tetragonal and orthorhombic FE phases. Monoclinic phases and their manifestation in properties of PbZr$_{1-x}$Ti$_x$O$_3$ and Pb(BI$_{1/3}$BII$_{2/3}$)O$_3$-PbTiO$_3$ solid solutions have been an object of intense studies.[8-10]

The monoclinic (M$_C$) phase in BaTiO$_3$ was studied in ref. [11-15]. High precision X-ray studies of (001) field cooled crystals[11] showed that this phase was stable below 300 K after removal of the electric field. Presence of the two-phase coherent hybrid crystal structure in single crystals of BaTiO$_3$ was observed by Raman spectra.[12] Studies of BaTiO$_3$ ceramics by the high-resolution synchrotron X-ray powder diffraction and atomic resolution aberration-corrected transmission electron microscopy, in conjunction with a powder poling technique[15] reveal that the equilibrium state of BaTiO$_3$ at 300 K is characterized by the coexistence of the metastable monoclinic, orthorhombic and tetragonal phases.

The goal of the present study is a theoretical analysis of the temperature intervals of phase stability and their coexistence in barium titanate. Our analysis uses a phenomenological approach based on the Ginzburg-Landau-Devonshire thermodynamic potential including terms up to the eighth power of the polarization vector components, which are necessary for description of the monoclinic phase. First, we present the analysis of the temperature intervals of stability for the cubic, tetragonal, orthorhombic, and monoclinic phases. Then, we present some results on the internal structure of interphase boundaries separating domains of coexisting phases within the temperature regions of phase coexistence near first order phase transitions.

The thermodynamic potential includes the parts describing the following subsystems: the polar ($\Phi_{Polar}$), the elastic ($\Phi_{Elastic}$), and the part responsible for the interaction between the polar and elastic subsystems due to electrostriction ($\Phi_{Striction}$).

$$\Phi(P_i, u_k) = \Phi_{Polar}(P_i) + \Phi_{Elastic}(u_{ij}) + \Phi_{Striction}(P_i, u_{ij}) \tag{1}$$

The polar part contains terms up to the eighth power of the polarization vector components as is necessary to describe the low symmetry phases.[16,17]

$$\begin{aligned}
\Phi_{Polar}(P_i) &= \alpha_1\left(P_1^2 + P_2^2 + P_3^2\right) + \alpha_{11}\left(P_1^4 + P_2^4 + P_3^4\right) + \alpha_{12}\left(P_1^2 P_2^2 + P_2^2 P_3^2 + P_1^2 P_3^2\right) \\
&+ \alpha_{111}\left(P_1^6 + P_2^6 + P_3^6\right) + \alpha_{112}\left[P_1^4\left(P_2^2 + P_3^2\right) + P_2^4\left(P_1^2 + P_3^2\right) + P_3^4\left(P_2^2 + P_1^2\right)\right] \\
&+ \alpha_{123} P_1^2 P_2^2 P_3^2 + \alpha_{1111}\left(P_1^8 + P_2^8 + P_3^8\right) + \alpha_{1122}\left(P_1^4 P_2^4 + P_2^4 P_3^4 + P_1^4 P_3^4\right) \\
&+ \alpha_{1112}\left[P_1^6\left(P_2^2 + P_3^2\right) + P_2^6\left(P_1^2 + P_3^2\right) + P_3^6\left(P_1^2 + P_2^2\right)\right] \\
&+ \alpha_{1123} P_1^2 P_2^2 P_3^2 \left(P_1^2 + P_2^2 + P_3^2\right) - \left(\vec{P} \cdot \vec{E}_{ext}\right)
\end{aligned} \tag{2}$$

$$\Phi_{Elastic}(u_{ij}) = \frac{1}{2}c_{11}\left(u_1^2 + u_2^2 + u_3^2\right) + c_{12}\left(u_1 u_2 + u_2 u_3 + u_1 u_3\right) + \frac{1}{2}c_{44}\left(u_4^2 + u_5^2 + u_6^2\right) \tag{3}$$

$$\begin{aligned}
\Phi_{Striction}(P_i, u_{ij}) &= -q_{11}\left(P_1^2 u_1 + P_2^2 u_2 + P_3^2 u_3\right) - q_{12}\left[P_1^2(u_2 + u_3) + P_2^2(u_1 + u_3) + P_3^2(u_1 + u_2)\right] \\
&- q_{44}\left(P_2 P_3 u_4 + P_1 P_3 u_5 + P_1 P_2 u_6\right)
\end{aligned} \tag{4}$$

In these formulas $P_i$ ($i = 1,2,3$) are the components of the polarization vector, $\vec{E}_{ext}$ is an external electric field vector, $u_\beta$ ($\beta = 1, 2, \ldots, 6$) are the components of the strain tensor in Voigt notation, $\alpha_1$, $\alpha_{ij}$, $\alpha_{ijk}$, $\alpha_{ijkl}$ are the phenomenological coefficients, $c_{ij}$ are elastic moduli, and $q_{ij}$ are parameters describing electrostriction interactions. Values for the coefficients appearing in (2-4) used for calculations are given in ref. [18].

The standard procedure of minimization of the thermodynamic potential has been described in detail.[4] We considered a single crystal of BaTiO$_3$ not subjected to external stress. The system of equations

$$\frac{\partial \Phi}{\partial P_i} = 0 \quad \text{and} \quad \frac{\partial \Phi}{\partial u_\beta} = 0 \tag{5}$$

give well known solutions (See Ref. [18-21] for example) for components of the polarization vector and components of the strain tensor in all homogeneous phases possible in BaTiO$_3$. Using these solutions, one can find the temperature dependence of components and magnitude of the homogeneous polarization at temperatures below $T_C$ and obtain the temperature dependencies for the thermodynamic potential.

We must note that inclusion of the eighth order terms in (2) does not lead to any significant differences in the results for the polarization vector and profiles of domain walls (DW) existing in the tetragonal and orthorhombic phases obtained within the sixth-order model.[21]

The temperature intervals of phase stability are determined from the condition that the determinant of the Hessian matrix of second derivatives of the thermodynamic potential with respect to all variables determining the phase must be a positive value.

$$\left|\frac{\partial^2 \Phi}{\partial \eta_i \partial \eta_j}\right| \geq 0, \tag{6}$$

Here $\eta_i$ constitutes three components of the polarization vector $P_i$ and six components of the strain tensor $u_\beta$. The second derivatives must be calculated with the substitution of solutions to equations (5) for each phase.

The temperature intervals of stability for the paraelectric, tetragonal, orthorhombic, and monoclinic homogeneous phases have been obtained. Near the phase transition from the paraelectric to the tetragonal FE phase, the phases coexist within the temperature interval $388 \leq T \leq 402\ °K$. This is nearly three times larger than the interval of the hysteresis behavior of the dielectric constant in the vicinity of the Curie point (~ 5 °C) observed in ref. [1,7]. Such difference is because the actual phase transition taking place in the crystal is the transition from the homogeneous paraelectric state into the inhomogeneous state of FE domains in the tetragonal phase. The same situations take place at the points of phase transitions from the tetragonal to the orthorhombic phase as well as from the orthorhombic to the rhombohedral phase since different types of domains are present in all these phases. Detailed analysis of the phase transition into inhomogeneous FE state with domains will be considered elsewhere. Our results on the stability regions for pure (single domain) phases in BaTiO$_3$ showing rather wide temperature intervals of phase coexistence agree with the studies on the influence of intrinsic interactions between domains of different phases in BaTiO$_3$.[22,23]

Our analysis shows that the monoclinic phase has a rather wide temperature interval where it remains stable. However, the minimum of the thermodynamic potential corresponding to this phase is located above the minima for tetragonal and orthorhombic FE phases within the whole temperature interval of the monoclinic phase existence. Thus, our findings resonate with previous conclusions[15] about the instability regime in BaTiO$_3$ at room temperature where domains of the monoclinic phase may appear as metastable regions.

The temperature dependence of the density of the thermodynamic potential for all phases appearing in BaTiO$_3$ is plotted in figure 1. The temperature intervals of phase stability are marked with gray dashed lines for each phase. The first order phase transition temperatures are marked by the following notations: $T_C$ is the Curie point, $T_{T\text{-}O}$ is the tetragonal-orthorhombic phase transition temperature and $T_{O\text{-}R}$ is the orthorhombic-rhombohedral phase transition temperature.

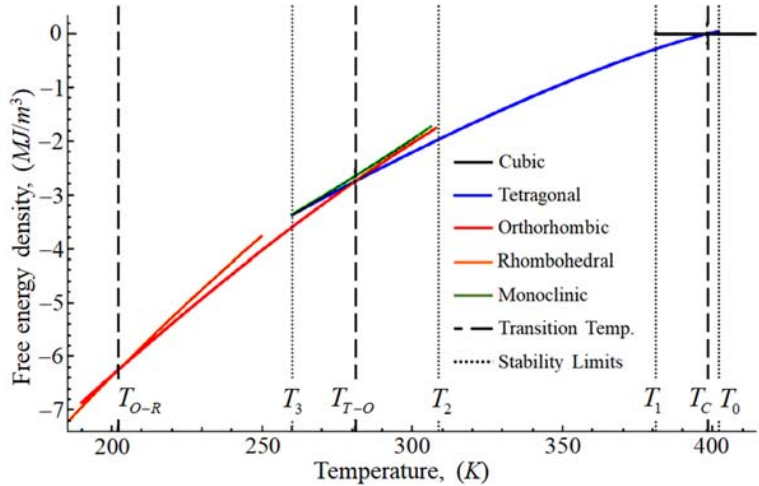

Figure 1: Temperature dependence of the thermodynamic potential for all phases appearing in BaTiO$_3$.

The plot is truncated at $T = 200\ K$, since only the rhombohedral phase occurs in the low temperature region. In figure 1, the interval $T_1 \leq T \leq T_0$ marks the temperature region within which the cubic and tetragonal phases coexist. The interval $T_3 \leq T \leq T_2$ corresponds to the region of coexistence of the tetragonal, orthorhombic, and monoclinic phases. The upper temperature stability limit for rhombohedral phase is $T_4$

= 325 *K*. The temperature $T_4$ is not shown in figure 1 and the energy density function of the rhombohedral phase is truncated for clarity. The orthorhombic-rhombohedral transition temperature is equal to 202 *K*.

Here we want to note several important points. As seen in fig. 1 the difference in energies between the monoclinic and tetragonal and orthorhombic phases is quite small, which is in agreement with generally noted small energy differences between FE phases[24,25] due to small values of distortions of the original pseudo-cubic perovskite crystal structure. The monoclinic phase is a metastable phase for the entire temperature interval shown in figure 1. It is interesting that the interval of existence of the monoclinic phase coincides with the interval of coexistence of the tetragonal and orthorhombic phases $T_3 \leq T \leq T_2$. Thus, the metastable domains of the monoclinic phase can be present in the tetragonal phase. There exist experimental observations of such domains present in the tetragonal phase.[26-28] However, the coexistence of the monoclinic and tetragonal domains in $BaTiO_3$ at room temperature was disputed by the authors of ref. [29] who stated that the actual twinning process can explain the observed diffraction pattern and it is not necessary to invoke a monoclinic symmetry.

Analysis of the polarization distribution for DWs separating domains appearing within stable FE phases has been carried out in several previous publications.[21,30-34] The only analysis of the polarization distribution within the interphase boundary (IPB) separating coexisting domains of cubic and tetragonal phases has already been considered in ref. [34] only for the case $T = T_C$. The detailed analysis of IPBs separating domains of coexisting tetragonal, orthorhombic, and monoclinic phases have not been carried out until now.

The polarization distributions, width, and surface energies for the IPBs separating domains of all coexisting phases in the vicinity of first-order phase transitions have been obtained. The influence of the higher order invariants in the thermodynamic potential on the structure of DWs separating domains within the tetragonal and orthorhombic FE phases that were studied earlier using thermodynamic potential that included lower order invariants have been considered.

Here we present our results on IPBs separating the tetragonal and monoclinic phases only. The more detailed presentation of results on IPBs between all coexisting phases and the influence of higher order invariants in thermodynamic potential on polarization distributions in DWs within tetragonal and orthorhombic phases will be presented elsewhere.

To analyze the polarization distribution within DWs and IPBs, the gradients of the polarization vector must be added to the thermodynamic potential (2). The gradient part commonly used for such analysis[31] is:

$$\Phi_{Grad} = \frac{g_{11}}{2}\left(P_{1,1}^2 + P_{2,2}^2 + P_{3,3}^2\right) + g_{12}\left(P_{1,1}P_{2,2} + P_{2,2}P_{3,3} + P_{1,1}P_{3,3}\right) \\ + \frac{g_{44}}{2}\left[\left(P_{1,2}+P_{2,1}\right)^2 + \left(P_{2,3}+P_{3,2}\right)^2 + \left(P_{1,3}+P_{3,1}\right)^2\right] \quad (7)$$

Here $g_{ij}$ is the isotropic gradient tensor (values of components of $g_{ij}$ were taken from ref. [31]), $P_{i,j} \equiv \partial P_i / \partial x_j$ where $i, j = 1, 2, 3$ and 1,2,3 stands for the principal Cartesian coordinates (*x*, *y*, *z*). The distribution of the polarization vector inside the IPBs follows from solutions of the Euler-Lagrange equations in which the variable *s* is the coordinate normal to the domain wall.

$$\frac{\partial}{\partial s}\frac{\partial \varphi}{\partial P_i'} - \frac{\partial \varphi}{\partial P_i} = 0 \qquad P_i' \equiv \frac{\partial P_i}{\partial s} \quad (8)$$

Based on results of ref. [13], the IPB between the monoclinic and tetragonal phases can be parallel to the (101) plane, normal to the *y*-axis. We call this type of IPB as parallel (∥ Tetragonal/Monoclinic IPB).

The monoclinic phase has two non-zero, non-equivalent components of the polarization vector oriented within the (101) plane $(P_1 \neq P_3 \neq 0)$. The region of the tetragonal domain corresponds to $s \to -\infty$ and the one of the monoclinic domain to $s \to +\infty$. The following boundary conditions were chosen to solve equations (8).

$$P_1(-\infty) = 0, \quad P_1(+\infty) = P_{1,0} \leq P_0/\sqrt{2} \quad \& \quad \left.\frac{dP_1}{ds}\right|_{s \to \pm\infty} = 0$$

$$P_3(-\infty) = P_0, \quad P_3(+\infty) = P_{3,0} \geq P_0/\sqrt{2} \quad \& \quad \left.\frac{dP_3}{ds}\right|_{s \to \pm\infty} = 0$$

The following notation is used for the equilibrium solution far from the interphase boundary: for the $M_C$ phase respective components of the polarization solution are denoted as $P_{1,0}$ and $P_{3,0}$ with the magnitude of the polarization denoted as simply $P_0 = \sqrt{P_{1,0}^2 + P_{3,0}^2}$. This geometry corresponds to the situation when the polarization vector in the domain of tetragonal phase is parallel to the plane of location of the polarization vector of the monoclinic phase and parallel to the interphase boundary.

The polarization solution for the tetragonal region far from the boundary is determined using the same eighth order thermodynamic potential. The interphase boundary separates the coexisting tetragonal domain (left) and the monoclinic domain (right). It was noted[6,13] that the monoclinic symmetry observed in BaTiO$_3$ is very close to orthorhombic; therefore, we used the set of parameters from ref. [18] for calculations. Dependency of the polarization vector components on the normal coordinate ($s$) for such tetragonal/monoclinic IPB (∥ tetragonal/monoclinic IPB) is presented in figure 2.

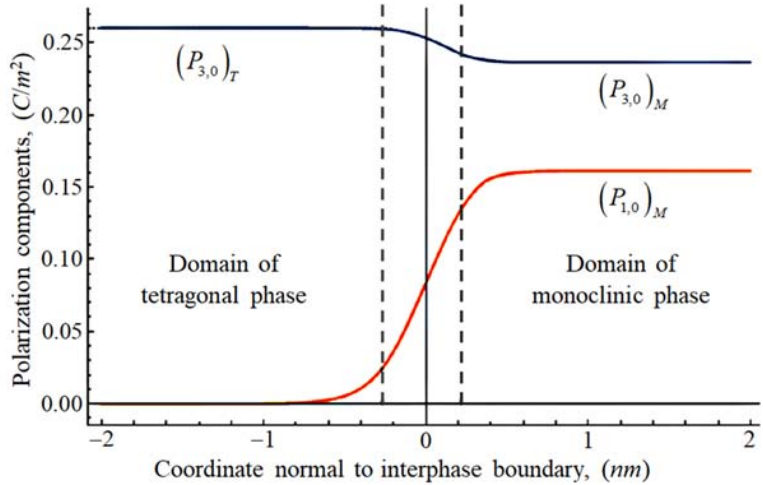

Fig. 2. Polarization distribution inside the ∥ tetragonal/monoclinic IPB. The effective thickness of the interphase boundary is marked by vertical dashed lines is given for reference. $P_3(s)$ – Blue, $P_1(s)$ – Orange.

The thickness of the IPB was defined in a way similar to the method[21,31] used for DWs and marked by the vertical dashed lines in figure 2. The changes of the polarization components with the coordinate normal to the ∥ tetragonal/monoclinic IPB happens within the plane of the boundary.

A different situation, hereafter referred to as the ⊥ tetragonal/monoclinic IPB is also possible. The polarization components inside this IPB depend on the normal coordinate in a similar manner to the theorized behavior of the polarization components inside the 90° DW; therefore, we used a rotated Cartesian coordinate system as it was done in ref. [32, 33] for analysis of the 90° DW. Two components of the polarization in rotated coordinate system are the component normal to the interphase boundary, $P_{R'}$ and the component parallel to the boundary, $P_S$. The boundary conditions used for solution of equations (8) in this case are as follows

$$P_1(-\infty) = 0, \qquad P_1(+\infty) = P_{1,0}$$
$$P_3(-\infty) = P_0, \qquad P_3(+\infty) = P_{3,0}$$

It has to be noted that the strain terms in the thermodynamic potential have been considered during the analysis of coordinate dependence of polarization components on coordinate normal to the IPB in the same way as it was done for 90° DW in ref. [32]. The dependency of these polarization components on the coordinate normal to the interphase boundary are presented in figure 3.

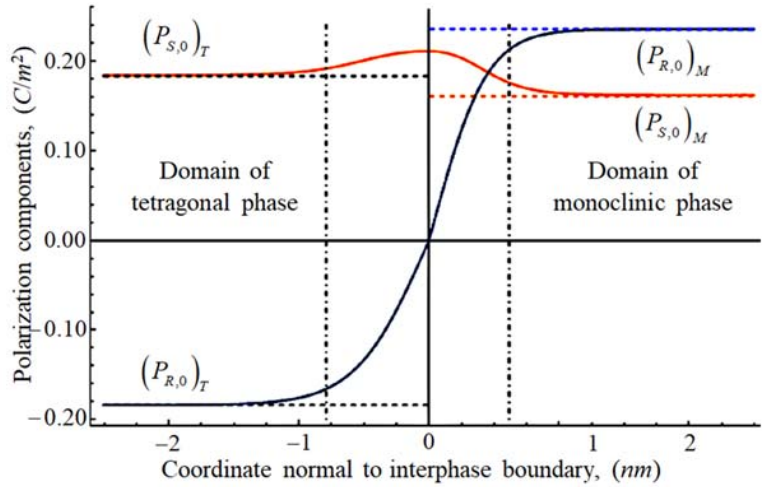

Figure 3. Components of the polarization distribution inside the ⊥ tetragonal/monoclinic IPB.

The same analysis has been done for IPBs separating the coexisting orthorhombic and monoclinic phases. Details of these calculations will be published elsewhere.

The next step in our consideration was to compare the effective thickness of the IPBs and compare obtained results to known thickness of DWs inside the phases. Our results on the temperature dependence of all IPBs are presented in figure 4.

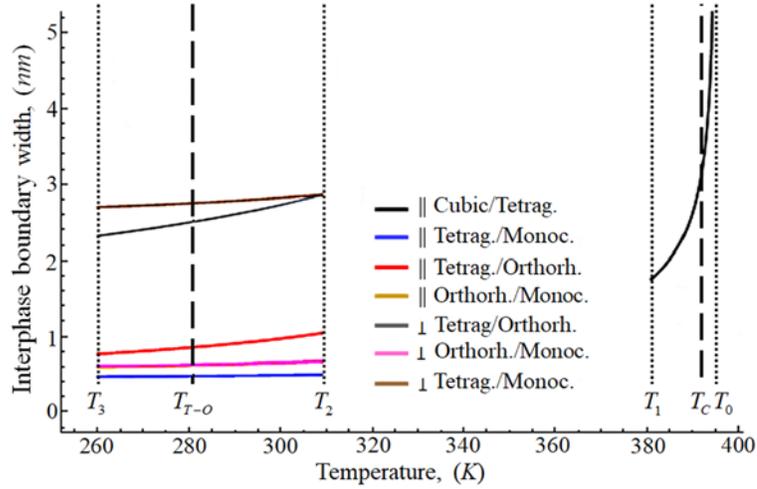

Figure 4. Temperature dependence of the thickness of the interphase boundaries.

The phase transition temperatures (black-vertical dashed lines) and the points of stability loss (vertical gray dotted lines) are given for reference in figure 4. It should be reminded that the interval of coexistence for the tetragonal, orthorhombic, and the metastable monoclinic phases is $T_3 \leq T \leq T_2$ (See also fig 1.).

      The values of the width of the IPBs at the room temperature are close to the known values of width of the DWs in tetragonal and orthorhombic phases.[21] However, the width of the ⊥ IPBs between tetragonal and monoclinic phases and the ⊥ IPB between the tetragonal and orthorhombic phases are larger than others due to the contributions of strains.

      We have also calculated the surface energy densities for all IPBs and compared them to the known values of energy densities of DWs possible within tetragonal and orthorhombic phases. The values of the energy densities for the IPBs depicted in figures 2 and 3 are 11.9 mJ/m$^2$ and 25.7 mJ/m$^2$, respectively. The values of the energy densities of 180° and 90° DW in the tetragonal phase[21,30] are 5.9 and 7 mJ/m$^2$. The energy density for the similar DW in the orthorhombic phase[21] is 8.4 mJ/m$^2$. Whereas, the energy of the IPB separating coexisting tetragonal and orthorhombic domains is 5.3 mJ/m$^2$ near room temperature.

      Based on these comparisons, one can conclude that the interphase boundaries between coexisting phases in the vicinity of first-order phase transitions possess a thickness close to the DW width occurring within phases, while having greater energy densities.